# Super- and Sub-critical Regions in Shocks driven by Radio-Loud and Radio-Quiet CMEs

*Bemporad A.[1] & Mancuso S.[1]

[1] **INAF - Osservatorio Astrofisico di Torino, via Osservatorio 20, 10025 Pino torinese (TO), Italy**

## Abstract

White-light coronagraphic images of Coronal Mass Ejections (CMEs) observed by SOHO/LASCO C2 have been used to estimate the density jump along the whole front of two CME-driven shocks. The two events are different in that the first one was a "radio-loud" fast CME, while the second one was a "radio quiet" slow CME. From the compression ratios inferred along the shock fronts, we estimated the Alfvén Mach numbers for the general case of an oblique shock. It turns out that the "radio-loud" CME shock is initially super-critical around the shock center, while later on the whole shock becomes sub-critical. On the contrary, the shock associated with the "radio-quiet" CME is sub-critical at all times. This suggests that CME-driven shocks could be efficient particle accelerators at the shock nose only at the initiation phases of the event, if and when the shock is super-critical, while at later times they lose their energy and the capability to accelerate high energetic particles.

## 1. Introduction

The last decades have seen a mounting interest of the scientific community in the study of the conditions at the Sun that can influence the performance of space-born and ground-based technological systems and that can affect human life and healt, namely the study of Space Weather. Our modern society became progressively vulnerable to disturbances associated with most powerful event occurring on the Sun, like solar flares (responsible for sudden terrestrial atmosphere heatings), Solar Energetic Particles (SEPs – which may damage satellite instrumentations and be dangerous for astronauts) and Coronal Mass Ejections (CMEs – responsible, among other effects, for geomagnetic storms).

In this regard, the formation of shock waves play an important role in the corona, because these waves are able to accelerate particles (electrons, protons, ions) up to near-relativistic energies. They are produced either as blast waves, due to the huge flare-induced pressure pulse, and/or piston-driven as bow shocks in front of fast Coronal Mass Ejections (CMEs). In the corona, they are detected in radio dynamic spectra, white-light images [1] and ultraviolet spectra [2, 3]. The shock represents a discontinuity with a transmitted mass flow, which is decelerated from super- to sub-Alfvénic speed [4]. It is thus a dissipative structure in which the kinetic and magnetic energy of a directed plasma flow is partly transferred to heating of the plasma. The dissipation does not take place, however, by means of particle collisions. Collisionless shocks can be divided into super- and sub-critical [5]:the critical fast Mach number $M_A^*$ is defined by equating the normal component of the downstream flow velocity in the shock frame to the sound speed. Supercritical shocks are important because usually produce much greater ion heating than subcritical shocks [6, 7]. In contrast to sub-critical shocks, resistivity in super-critical shocks cannot provide all the necessary dissipation for a shock transition according to the Rankine-Hugoniot relationships. Thus, other processes like wave-particle interactions provide the dissipation required for supercritical shock formation. This is the reason why they are able to accelerate SEPs efficiently to high energies. The SEP acceleration efficiency also depends on the angle $\theta_{Bn}$ between the magnetic field and the normal to the shock surface. In fact, the expansion of the CME fronts likely induces the formation of both quasi-parallel (i.e. $\theta_{Bn} \sim 0°$) and quasi-perpendicular (i.e. $\theta_{Bn} \sim 90°$) shocks, at the nose of the CME front and at the CME flanks, respectively [8]. Because the ion acceleration rate is faster in perpendicular than in parallel shocks, it is believed that SEPs are mostly accelerated in perpendicular shocks [9, 10]. Both kinds of shocks reflect ions, but in quasi-parallel shocks the combined geometries of the upstream field and of the typically curved shock surface is such that the reflected particles are enabled to escape upstream from the shock along the magnetic field. Hence, more in general, both quasi-parallel and quasi-perpendicular SEP accelerations are possible in CME-driven shocks.

Propagation of shocks in the solar corona and interplanetary medium is inferred from the detection of type II radio bursts (appearing as emission slowly drifting from high to low frequencies in dynamic radio spectra) which provide a direct radio signature of shocks [11]. Because every large SEP event is associated with a type-II burst, the latters are usually identified as strong indicators of particle acceleration by shocks. Usually, shocks producing a type-II burst are said to be "radio-loud" (RL), while those not producing a type-II burst are said "radio-quiet" (RQ), and the same terminology is applied to associated CMEs [12], even if this terminology is not fully correct because CMEs can be in general associated or not also with other kinds of radio emissions, like type-III and type-IV radio bursts [13, 14]. Statistical studies [15] demonstrate that RL CMEs are faster, wider and associated with stronger X-ray flares, but slow (v << 900 km/s) RL-CMEs and fast (v >> 900 km/s) RQ-CMEs are also observed, suggesting that conditions of the ambient corona (and in particular the local value of the Alfvén speed $v_A$) likely play a fundamental role in deciding the CME capability to accelerate shocks.

Thanks also to the availability of data acquired by STEREO spacecraft, [16] demonstrate that the type-II bursts (hence the CME-driven shocks) form when the CMEs are located at an heliocentric distance of ~ 1.5 solar radii, while weak or no shocks are observed around ~ 3-4 solar radii and that type II burst seems to end when the shock becomes subcritical. Hence, these results are in agreement with the idea that type-II bursts could be excited where the speed of the CME piston-driven shock exceed the local fast magnetosonic speed, which is expected to have a local minimum around 1.2-1.4 solar radii and a local maximum around 3.5 solar radii [17, 18]. Nevertheless, the exact location in the corona where the super- and sub-critical shock forms and how they evolve is at present unknown. In this work we extend our previous identification of super- and sub-critical regions along shock fronts observed in white light coronagraphic images [19] by focusing on two CME-driven shocks: the first event was a RL fast CME, while the second one was a RQ slow CME. As we are going to show here, the formation or not of type-II bursts can be associated with the presence or not of a super-critical region at the "nose" (i.e. center) of the shock. Data analysis and results are described in the next Section (§ 2) and discussed in the last Section (§ 3).

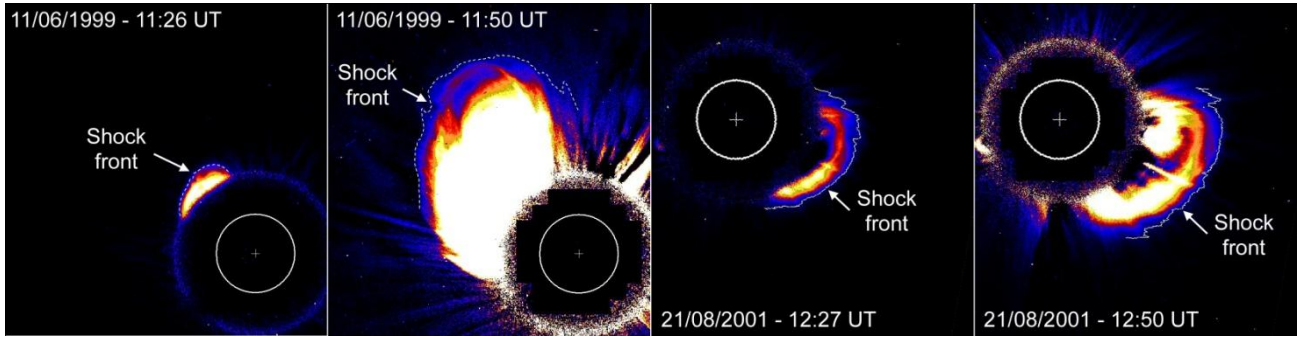

*Figure 1: Base difference LASCO/C2 images showing the location of the CME-driven shock front (dashed lines) for the radio-loud CME at 11:26 UT (left) and 11:50 UT (middle left) and for the radio-quiet CME at 12:27 UT (middle right) and 12:50 UT (right).*

## 2. Methodology and Results

The two events studied in this work are shown in Fig. 1 (top) as white light images acquired by the SOHO/LASCO-C2 coronagraph. In particular, this Figure shows a sequence of base difference images obtained by subtracting the intensity of the pre-CME corona to the CME images. The RL-CME, which occurred on 1999 June 11, was a fast event (propagating at a projected velocity of 1570 km/s) associated with a type II radio burst (detected by WIND/WAVES) and a C8.8 class flare (detected by GOES). On the contrary, the RQ-CME, which occurred on 2001 August 21, was a slow event (propagating at a projected velocity of 540 km/s) without radio burst and without flare. White light images have been employed first to derive the pre-CME coronal electron densities $n_e$ (cm$^{-3}$): a good knowledge of the ambient corona electron density is important in order to estimate the shock compression ratio from the ratio between the white light intensities observed at the shock front and in the unperturbed corona.

Densities have been derived from polarized Brightness (pB) images acquired by LASCO before each event: in particular LASCO/C2 instrument acquired the last pB images before each CME on June 10, 1999 at 21:00 UT and on August 20, 2001 at 21:00UT; no significant changes in the white light corona occurred between these times and the occurrence of the CMEs. The pB images have been analyzed with standard inversion routine provided within the SolarSoftware (pb_inverter.pro) which assumes spherical symmetry to perform the classical Van Der Hulst inversion, obtaining a set of coronal electron density radial profiles all over the region of the shock propagation with an angular resolution by 3 degrees.

Second, we identified the location of the shock fronts (dashed lines in Fig. 1) at different latitudes as recently done by [20], i.e. by extracting radials in each base difference image at different latitudes and by automatically identifying the location of the white-light intensity increase located above the expanding CME front. Third, we estimated, from the ratio between the white light intensities observed at the shock front and in the unperturbed corona, the shock compression ratio $X = \rho_d/\rho_u$ between the downstream ($\rho_d$) and upstream ($\rho_u$) densities (Fig. 2, top). Line-of-sight integration effects have been also taken into account in the determination of $X$ [3, 19]. As in [3] the shock compression ratios have been estimated by assuming constant values at different latitudes for the shock depth $L$ along the line of sight, which have been estimated from the 2-D projected thickness $d$ of the white light intensity increase across the shock (typically around $\sim 5 \times 10^4$ km) and by assuming that in 3-D the shock surface has the shape of an hemispherical shell. In the hypothesis of a plasma $\beta \ll 1$ ($\beta$ is the ratio between the thermal and magnetic plasma pressures), the shock Mach number $M_A$ (i.e., the ratio of the upstream flow speed along the shock normal to the upstream Alfvén speed) can be estimated from the compression ratio $X$ as

$$M_{A\perp} = \sqrt{\frac{X(X+5)}{2(4-X)}} \quad , \quad M_{A\parallel} = \sqrt{X} \quad , \quad M_{A\angle} = \sqrt{(M_{A\perp}\sin\theta_{Bn})^2 + (M_{A\parallel}\cos\theta_{Bn})^2},$$

where $M_{A\perp}$ ($M_{A\parallel}$) is the Mach number for perpendicular (parallel) shock, and the latter formula gives an order of magnitude estimate of the Mach number $M_{A\angle}$ for the general case of an oblique shock at the angle $\theta_{Bn}$.

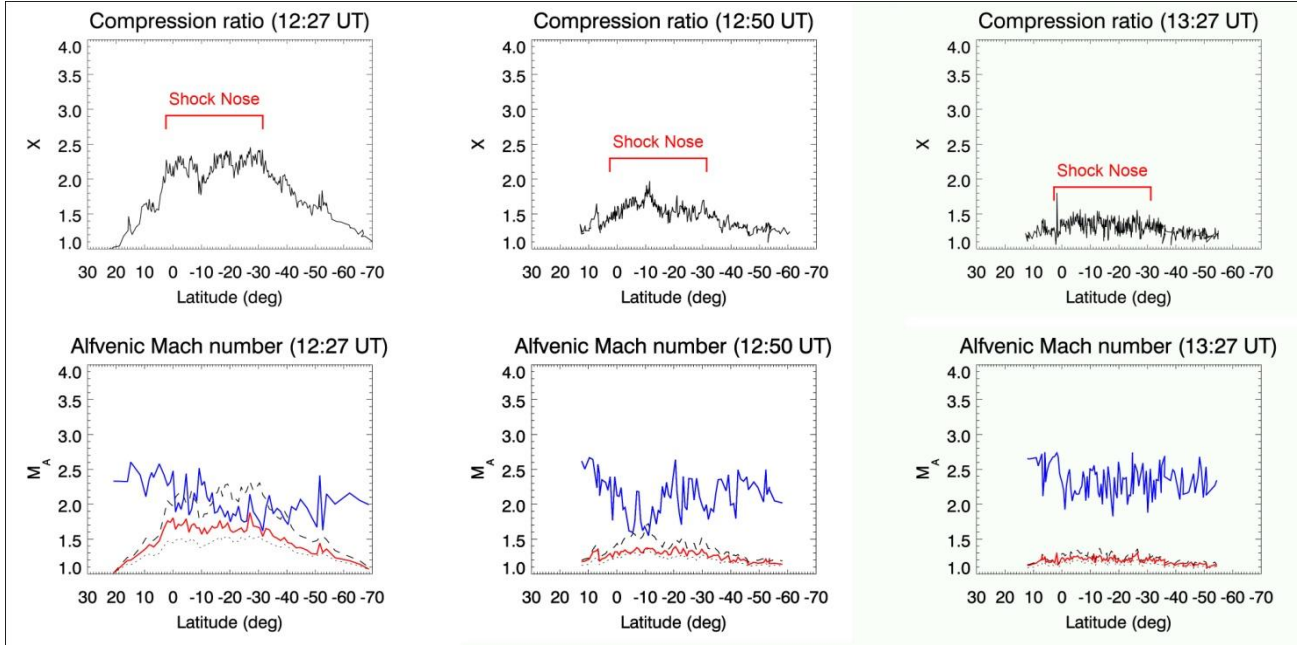

*Figure 2, top: compression ratios X = $\rho_d/\rho_u$ as measured at the points illustrated by the dashed lines in Figure 1 along the shock front of the radio-quiet CME at three different times. Bottom: theoretical Alfvén Mach numbers $M_A$ for perpendicular (dashed line) and parallel (dotted line) and for angles measured along the actual shock fronts (solid red lines) at 12:27 UT (left), 12:50 UT (middle) and 13:27 UT (right); the latter curves are compared with the corresponding critical Mach numbers (solid blue lines).*

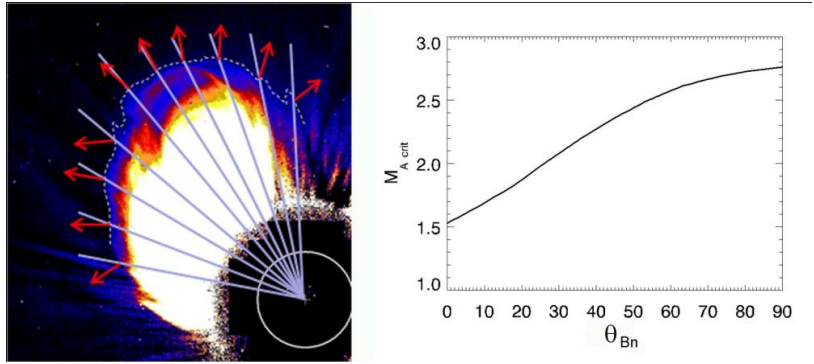

*Figure 3, left: cartoon showing how the angle $\theta_{Bn}$ between the magnetic field, assumed to be radial (cyan solid lines), and the shock normal (red arrows) has been derived along the shock front. Right: theoretical dependence of the critical Mach number $M_A^*$ as a function of $\theta_{Bn}$ in the limit of $\beta \ll 1$ [9].*

Given the quantities X (Fig. 2, top) and $\theta_{Bn}$ (Fig. 3, left) all along the shock front, Mach numbers $M_{A\parallel}$, $M_{A\angle}$ and $M_{A\perp}$ can be determined. Moreover, in the hypothesis of $\beta \ll 1$, the quantity $M_A^*$ is a monotonic function of $\theta_{Bn}$ (Fig. 3, right), hence can also be determined all along the shock front.

Resulting $M_{A\parallel}$, $M_{A\angle}$, $M_{A\perp}$ curves show in general a maximum at the latitudes corresponding to the shock center (or "nose"), while $M_A^*$ has a minimum at the same latitudes (Fig. 2, bottom). Interestingly, for the RL-CME we found $M_{A\angle} > M_A^*$ only at the early stages (11:26 UT) and $M_{A\angle} < M_A^*$ later on (11:50 UT – see Bemporad & Mancuso 2011, Fig. 2), while for the RQ-CME we found $M_{A\angle} < M_A^*$ at any time (Fig. 2, bottom). Hence, the shock super- or sub-criticality seems to be directly connected with the presence of a type II radio burst and likely of accelerated particles.

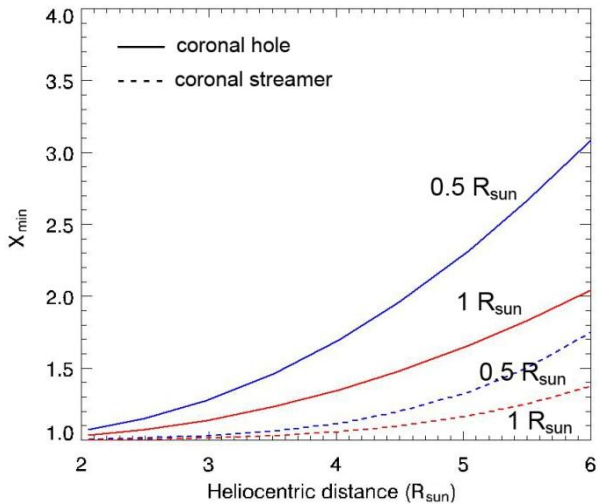

*Figure 4: typical values of the minimum shock compression ratio $X_{min}$ required for a $3\sigma$ detection of the shock in LASCO/C2 data at different altitudes. Values of $X_{min}$ have been computed by assuming shock thicknesses of $L = 0.5\ R_{sun}$ (blue lines) and $L = 1\ R_{sun}$ (red lines) for a shock observed in a typical coronal hole (solid lines) and a coronal streamer (dashed lines).*

Before concluding this Section, we want to show that the detection of interplanetary MHD shocks in LASCO white light images is possible in general, even if the expected density compressions are relatively small. This can be demonstrated as follows: if the depth of the shocked coronal region intercepted by the line of sight at a single pixel is $L$ (cm), then in order to detect within $3\sigma$ (cm$^{-2}$) the change in the column density across the shock, it is required that $L\rho_d - L\rho_u = L\rho_u (X-1) \geq 3\sigma$. This condition corresponds to a minimum shock strength $X_{min}$ for a $3\sigma$ detection given by $X_{min} = 1 + 3\sigma / L\rho_u$. Values of $X_{min}$ shown in Fig. 4 are computed with actual LASCO data at different altitudes by assuming thicknesses of $L = 0.5\ R_{sun}$ (blue lines) and $L = 1\ R_{sun}$ (red lines) for a shock observed in a typical coronal hole (solid lines) and a coronal streamer (dashed lines). This Figure shows that the minimum compression ratio $X_{min}$ required for shock detection in the white light LASCO images is in general well below the upper limit $X_{max} = 4$ in the LASCO/C2 field of view (2 – 6 $R_{sun}$), making the shock detectable, in general.

## 3. Discussion and Conclusions

As mentioned in the Introduction, recent observations show that: 1) statistically, RQ (RL) CMEs are slower (faster) and associated with weaker (stronger) flares [15]; 2) CME-driven shocks seem to be most efficient in accelerating electrons in the heliocentric distance range of 1.5 $R_s$ to 4 $R_s$ [16];

3) RQ shocks are likely subcritical, whereas RL shocks are supercritical [21]; 4) the Alfvénic Mach numbers of shocks with a SEP event are on average 1.6 times higher than those of shocks without [22]; 5) there is very close association between the CME nose and the 1st type II burst and between the CME-streamer interaction and the 2nd type II burst [23]. In agreement also with these results, our study suggests that:

1. type-II radio bursts (associated with the propagation of CME-driven shocks) are likely produced where the shock is strong enough to be supercritical (red region in Fig. 5);
2. the supercritical region is located at the shock "nose", where quasi-parallel shock occurs;
3. as the shock propagates, it slows down, the supercritical region disappears, and the whole shock becomes subcritical (blue region in Fig. 5).

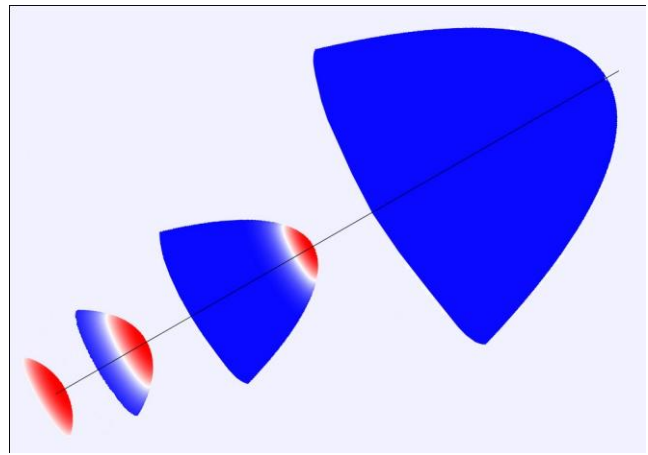

*Figure 5: schematic showing the possible evolution of supercritical (red) and subcritical (blue) regions over the shock surface.*

This indicates that particle acceleration likely occurs at the quasi-parallel shock only where and when the shock is strong enough to be supercritical. These results are in very good agreement for instance with those obtained by [24], who demonstrate that "solar type II radio bursts should be considered to be generated either by weak supercritical, quasi-parallel, or by subcritical, quasi-perpendicular fast magnetosonic shock waves in the corona.". In agreement with the picture that type-II burst are produced by supercritical quasi-parallel shocks (as we concluded here), [25] proposed an electron acceleration model where short large amplitude magnetic field structures (SLAMS) detected in situ in quasi-parallel collisionless shocks may act as strong magnetic mirrors accelerating thermal electrons by multiple reflections. Hence, results presented here have potentially very important implications on the localization of particle acceleration and radio burst production sites and in the context of predictive space weather studies.

## Acknowledgements

A. B. acknowledges support from the European Commissions Seventh Framework Programme (FP7/2007-2013) under the grant agreement SWIFF (project no. 263340, www.swiff.eu).